\documentclass[reprint,eqsecnum,floats,aps,amsmath,amssymb,nofootinbib,prd,onecolumn, showpacs]{revtex4-2}

\usepackage{graphicx}
\usepackage{amsmath,amssymb}
\usepackage{hyperref}
\usepackage{mathrsfs}

\begin{document}
	
\title{Fock quantization of a Klein-Gordon field in the interior geometry of a nonrotating black hole}
	
\author{Jer\'onimo Cortez}
\email{Electronic address: jacq@ciencias.unam.mx}
\affiliation{Departamento de F\'{\i}sica, Facultad de Ciencias, Universidad Nacional Aut\'onoma de M\'exico, Ciudad de M\'exico 04510, Mexico}
	
\author{Beatriz Elizaga Navascu\'es}
\email{Electronic address: bnavascues@lsu.edu}
\affiliation{Department of Physics and Astronomy, Louisiana State University, Baton Rouge, LA 70803-4001, USA}
	
\author{Guillermo A. Mena Marug\'an}
\email{Electronic address: mena@iem.cfmac.csic.es}
\affiliation{Instituto de Estructura de la Materia, IEM-CSIC, C/ Serrano 121, 28006 Madrid, Spain}
	
\author{Alvaro Torres-Caballeros}
\email{Electronic address: alvaro.torres@iem.cfmac.csic.es}
\affiliation{Instituto de Estructura de la Materia, IEM-CSIC, C/ Serrano 121, 28006 Madrid, Spain}
	
\author{Jos\'e M. Velhinho}
\email{Electronic address: jvelhi@ubi.pt}
\affiliation{Faculdade de Ci\^encias and FibEnTech-UBI, Universidade da Beira Interior, R. Marqu\^es D'\'Avila e Bolama, 6201-001 Covilh\~a, Portugal}
	
\begin{abstract}
We study the canonical quantization of a scalar field in a Kantowski-Sachs spacetime. For simplicity, we consider compactified spatial sections, since this does not affect the ultraviolet behavior. A time-dependent canonical transformation is performed prior to quantization. As in previously studied cases, the purpose of this canonical transformation is to identify and extract the background contribution to the field evolution which is obstructing a unitary implementation of the field dynamics at the quantum level. This splitting of the time dependence into a background piece and the part to be seen as true quantum evolution is to a large extent determined by the unitarity requirement itself. The quantization is performed in the usual setup of Fock representations, demanding the preservation of the spatial symmetries. Under the joint requirements of quantum unitary dynamics and compatibility with those classical symmetries, the quantization is shown to be unique, in the sense that any two representations with these properties are unitarily equivalent. This confirms the validity of our conditions as criteria to discriminate among possibly inequivalent quantum descriptions. The interest of this analysis goes beyond cosmological applications since the interior of a nonrotating black hole has a geometry of the Kantowski-Sachs type.
		
\end{abstract}

\maketitle
	
\section{Introduction}
\label{int}
	
The class of Kantowski-Sachs gravitational models describe spherically symmetric homogeneous spacetimes, with a spatial manifold of the form $\mathbb{R} \times \mathbb{S}^{2}$ \cite{KS1,KS2,KS3}. Anisotropy is introduced by allowing the metric on the two-sphere $\mathbb{S}^{2}$ to evolve independently of the radial metric component. Together with other known anisotropic models, Kantowski-Sachs models therefore have intrinsic interest in cosmological studies, namely as testing grounds to analyze possible anisotropy effects. In particular, the study of quantum fields in Kantowski-Sachs spacetimes is well motivated, both to analyze the propagation of matter fields in a homogeneous scenario and to study deviations from homogeneity. 
	
On the other hand, it is well known that the interior geometry of a nonrotating, uncharged black hole can be described by a Kantowski-Sachs metric \cite{AB}. In fact, the temporal and radial coordinates switch roles in the interior of a Schwarzschild black hole, and we are left with a natural foliation by spatial manifolds of the form $\mathbb{R} \times \mathbb{S}^{2}$ and a spatially homogeneous metric that, in suitable coordinates, is precisely of the Kantowski–Sachs type. 
	
In this respect, the study of quantum matter fields in the Kantowski-Sachs context is interesting also from the  perspective of black hole physics. Furthermore, quantum scalar fields in Kantowski-Sachs spacetimes are expected to be a component of more comprehensive approaches to black hole physics, namely in order to include matter or perturbations in full quantum models for black holes. 
	
In particular, quantum models combining Loop Quantum Gravity methods \cite{ALQG} with Fock quantization techniques (for the treatment of local degrees of freedom), require a well defined, rigorous, and unambiguous quantization of (test) fields in Kantowski-Sachs spacetimes. The interest in developing these models has recently been triggered by the proposal of an effective extension in Loop Quantum Cosmology of the Schwarzschild-Kruskal spacetime, with an interior region that is isometric to a Kantowski-Sachs cosmology \cite{AOS, AOS2, AO}. This proposal includes techniques for determining the exterior geometry. The possibility of introducing quantum perturbations around a loop quantum version of this black hole model \cite{EGM,EMM}, dealing with the composed system in the framework of the so-called hybrid Loop Quantum Cosmology approach \cite{hLQC,FMO,CMM}, opens new avenues to the study of quantum effects on the gravitational radiation in the asymptotic exterior regions.
	
In the implementation of this strategy, one must face the well-known ambiguities in the quantum description of field theories, which appear even in the context of the usual Fock representations of the canonical commutation relations (CCR) \cite{wald}. In fact, the standard Schr\"odinger quantization (and unitarily equivalent versions thereof) is only available for a finite number of degrees of freedom. In quantum field theory (QFT), on the other hand,  there are in principle infinitely many nonequivalent representations of the CCR. Restricting attention to linear theories and representations of the Fock type, this ambiguity is parametrized by the class of complex structures in the space of classical solutions. Equivalently, each representation corresponds to a choice of creation and annihilation pairs, or to the selection of the corresponding vacuum, in more physical terms. In this context, the selection of a proper and uniquely defined Fock quantization typically relies on the requirement of invariance under spacetime symmetries. A key role is played by time-translation invariance \cite{kayb}, which however is only available in the special case of stationary spacetimes.
	
To cope with the ambiguity in the Fock quantization of linear fields in nonstationary scenarios, a programme has been ongoing \cite{ccmv,cmov2,cmov3,cmov4,fmov,CJCAP,cmv,cmv2}, aiming at the establishment of natural criteria that, when fulfilled, could deliver a unique quantization (or rather a unique class of equivalent quantizations). The proposed criteria are actually quite simple and well motivated, consisting of the two following requirements: \emph{i\/)} invariance under the available spatial symmetries; \emph{ii\/)} unitary quantum dynamics, meaning a unitary implementation of the set of linear canonical transformations defined by classical evolution. Note that the second requirement is the natural replacement of the unavailable stronger requirement of invariance under time translations. In particular, it leads to a well-defined Schr\"odinger picture with a standard probabilistic interpretation, which could be especially important in the context of quantum fields in black hole spacetimes. 
	
The viability and the effectiveness of the proposed criteria to select a unique quantization were proven in a variety of situations \cite{ccmv,fmov,cmv2,cmvT,cmsv,CM,cmvS}, including Klein-Gordon fields in homogeneous and isotropic cosmologies \cite{cmov2,CJCAP}. It turns out that, in order to achieve unitary dynamics in the first place, a time-dependent redefinition of the field and momentum canonical pair is necessary, which is determined by the time evolution of the spacetime background itself. While in a standard physical situation, not involving gravitational degrees of freedom, a time-dependent canonical transformation prior to quantization could be seen as problematic, one should keep in mind that such transformations are in fact common practice in the cosmological context. In fact, for perturbation theory in isotropic cosmological scenarios, the above-mentioned field redefinition corresponds simply to the well-known Mukhanov-Sasaki variables. Furthermore, in every case where such a redefinition was seen to be necessary, it was also shown to be essentially unique, in the sense that no other time-dependent transformation could lead to a different quantization with unitary dynamics \cite{cmv2}. Thus, no new ambiguity is introduced here. On the one hand, the insistence on unitary dynamics requires a splitting of the full time dependence of fields in nonstationary backgrounds. On the other hand, this splitting is fully and uniquely specified by natural requirements. Given the status of time in General Relativity, and the thus far illusive full compatibility between Quantum Mechanics and General Relativity, it seems advisable to keep an open mind and to let the formalism itself guide us in selecting the formulation which gives the most amenable mathematical-physics output.   
	
Most of the situations where the proposed criteria have been put to the test are conformally ultrastatic \cite{cmov4}, such as scalar fields in homogeneous and isotropic cosmological models \cite{cmv,cmv2}. In these cases, the time-dependent canonical transformations that give way to unitary dynamics are local, essentially field rescalings. Nonetheless, this framework has been generalized to other models where conformal transformations play no relevant role. An example are massive fermionic fields in cosmological spacetimes. For them, the canonical transformation is not the same for different helicities and chiralities \cite{fermi,fermi2,fermi3,fermi4}. More important for our discussion is the case of a Klein-Gordon field in a Bianchi I universe \cite{BI}. To tame the anisotropy of this cosmological model, one has to consider canonical transformations that are nonlocal inasmuch as they vary for different modes of the spatial Laplace-Beltrami operator (see also Ref. \cite{fon} for another model where these methods have been applied). If one allows for mode-dependent canonical transformations, the requirements of invariance under the spatial symmetries and unitary dynamics indeed determine a unique possibility to construct a viable Fock quantization.
	
In the present work, we extend our analysis to the case of a Klein-Gordon field in another anisotropic cosmology: this time in a Kantowski–Sachs spacetime. The interest of this analysis goes beyond proving that the proposed criteria remain valid in more general anisotropic scenarios. We also have in mind the potential applications to black hole physics (see e.g. \cite{AOS,KSBH1,KSBH2,KSBH3,KSBH4,KSBH5}). The uniqueness of the Fock quantization of fields, the unitarity of their dynamics, and the equivalence of their vacua in models for the interior geometry of black holes are important preliminary steps to address outstanding issues of quantum field theory in their exterior.
	
This work is organized as follows. In Sec.\ \ref{sec:Hamil} we present the Hamiltonian analysis for a scalar field in a   Kantowski–Sachs geometry, including its natural mode decomposition. For mathematical convenience, the radial coordinate in the spatial manifold is compactified in a circle and the field is considered to be massless, although in this context a nonzero value of the mass would bring no further complication. In Sec.\ \ref{ncv} a time-dependent and mode-dependent canonical transformation is presented, as a necessary step to achieve unitary dynamics at the quantum level. The classical dynamics of the new variables introduced in this way is analyzed in Sec.\ \ref{acd}, with particular emphasis on the ultraviolet region. Considerations of unitary equivalence of representations and unitary implementation of classical canonical transformations, which permeate our whole analysis, depend exclusively on the ultraviolet behavior, and not on any finite number of modes. Our results are presented in Secs. \ref{fock} and \ref{uni}. First, we consider the class of Fock representations which respect the spatial symmetries (and hence the corresponding symmetries of the dynamical mode equations) and we derive the conditions imposed on them by the requirement of unitary dynamics, thus selecting our privileged subclass of representations. Different Fock representations are here characterized by a particular definition of the variables to be promoted to creation and annihilation operators. We include a time dependence in this definition, which simultaneously allows us to show that the dynamics put forward by the canonical transformation of Sec.\  \ref{ncv} is fixed by our criteria, up to  subdominant terms in the ultraviolet. Finally, in Sec.\ \ref{uni} we show that the quantization following from the fulfillment of our criteria is unique, in the sense that all the Fock representations within our selected subclass are unitarily equivalent. In Sec. \ref{con} we review our results and conclude.
	
\section{Klein-Gordon field in Kantowski-Sachs geometry: Hamiltonian formalism}
\label{sec:Hamil}
	
We consider a scalar field $\phi$ minimally coupled to a  Kantoswki-Sachs geometry. The Lagrangian density is of the form
\begin{equation} \label{1.1}
\mathcal{L}=   - \frac{1}{2} \sqrt{-g} g^{\mu \nu} \nabla_\mu \phi \nabla_\nu \phi ,
\end{equation}
where $g$ is the determinant of the metric and $g^{\mu \nu}$ is the inverse of the Kantowski-Sachs metric, defined as \cite{KS-SF}
\begin{equation}
\begin{aligned}
g_{\mu \nu}&=\text{diag}\left( -A^2(t),P^2(t),Q^2(t),Q^2(t) \ \text{sin}^2\theta\right), \\
ds^2 &= -A^2(t)dt^2 + P^2(t)dr^2 + Q^2(t) (d\theta^2 + \sin^2\theta \ d\varphi^2).
\end{aligned}
\end{equation}
	
To avoid unnecessary complications in the infrared limit, let us compactify the radial coordinate  $r$ in a circle of period $2\pi L_0$. One can always take $L_0$ sufficiently large and consider the noncompact limit. The topology of the spatial manifold is thus effectively $S^1 \times \mathbb{S}^{2}$. The variables $\theta$ and $\varphi$ are the standard coordinates on the sphere $\mathbb{S}^{2}$.
	
The corresponding Lagrangian $\mathbf{L}$, and therefore the Klein-Gordon equation for the field $\phi$, is written in terms of  the Laplace-Beltrami (LB) operator $\Delta$ associated with the spatial metric, which reads (at any given value of the time $t$)
\begin{equation}
\begin{aligned}
\Delta&= \frac{1}{P^2} \partial_r^2 + \frac{1}{Q^2}\left(  \partial^2_\theta + \cot\theta \partial_\theta  + \frac{1}{ \sin^2\theta} \partial_\varphi^2 \right) .
\end{aligned}
\end{equation}
We obtain
\begin{equation} \label{0.3}
\begin{aligned}
\mathbf{L} &= \int \mathcal{L}  dr  d\theta d\varphi = \int \frac{APQ^2}{2} \left( \frac{\dot{\phi}^2}{A^2} -h^{ij} \partial_i \phi \partial_j \phi \right) dr  d\Omega\\
&= \int \frac{APQ^2}{2} \left( \frac{\dot{\phi}^2}{A^2} + \phi \Delta\phi \right) dr d\Omega,
\end{aligned}
\end{equation}
where $d\Omega = \sin\theta d\theta d\varphi$ and integration by parts is involved in the last equality.
	
A basis of real eigenfunctions of the LB operator $\Delta$ is given by $s_{nlm}= F_n(r) \bar{Y}_l^m (\theta, \varphi) $, namely products of Fourier modes and real spherical harmonics, where the Fourier modes can be chosen as
\begin{eqnarray}
F_0(r)&=&\frac{1}{\sqrt{2\pi L_0}},\nonumber\\ 
F_n(r) &=& \frac{1}{\sqrt{\pi L_0}} \sin\left(\frac{nr}{L_0}\right)  \quad {\rm for} \quad n<0, \nonumber\\ 
F_n(r) &=& \frac{1}{\sqrt{\pi L_0}} \cos\left(\frac{nr}{L_0}\right) \quad {\rm for} \quad n>0,
\end{eqnarray}
and the real spherical harmonics as
\begin{eqnarray}
\bar{Y}_l^0 (\theta, \varphi) &=&Y_l^0 (\theta, \varphi), \nonumber\\
\bar{Y}_l^m (\theta, \varphi) &= &\frac{Y_l^m(\theta, \varphi)  +(-1)^mY_l^{-m}(\theta, \varphi) }{\sqrt{2}}, \quad {\rm for} \quad m>0, \nonumber\\
\bar{Y}_l^m (\theta, \varphi) &=& i \frac{Y_l^m(\theta, \varphi)  -(-1)^mY_l^{-m}(\theta, \varphi) }{\sqrt{2}}, \quad {\rm for} \quad m<0.
\end{eqnarray} 
Here, $Y_l^m$ are the complex spherical harmonics:
\begin{equation}
Y^m_l(\theta, \varphi) = (-1)^{m} \sqrt{\frac{(2l+1)}{4\pi} \frac{(l-m)!}{(l+m)!}} P^{m}_l(\cos(\theta)) e^{im\varphi},
\end{equation}
with $P^{m}_l$ being the Legendre polynomials, and we have used that $(Y_l^m)^{*}=(-1)^m Y_l^{-m}$, the symbol $^*$ denoting complex conjugation. For this basis, the corresponding eigenvalues (at time $t$) are minus
\begin{equation} \label{defW}
\begin{aligned}
W_{nl}(t) &= \frac{n^2}{P^2(t)}+ \frac{l(l+1)}{Q^2(t)} .
\end{aligned}
\end{equation}

The Klein-Gordon field can then be expanded in terms of the eigenfunctions of the LB operator, introducing a discrete set of dynamical modes $\{\phi_{nlm}(t)\}$ as follows:
\begin{equation}\label{defphi}
\phi(t,r,\theta,\varphi) =  \sum_{nlm} \phi_{nlm}(t)  F_n(r) \bar{Y}^m_l(\theta,\varphi) .
\end{equation}
Here, the label $n$ is summed over all integers, $l$ over all positive integers, and $m$ over the integers in the interval $[-l, l]$ for each given $l$. We notice that the Klein-Gordon field is real provided that its modes are real. 
	
From Eq. (\ref{0.3}), one can check that  the Klein-Gordon equation for the  field $\phi$ is equivalent to the following set of equations for the modes $\phi_{nlm}$:
\begin{equation}
\frac{1}{A^2} \ddot{\phi}_{nlm} +\left[ -\frac{\dot{A}}{A^3} + \frac{\dot{P}}{A^2P}+ \frac{2\dot{Q}}{A^2Q} \right] \dot{\phi}_{nlm} = W_{nl}(t)  \phi_{nlm} .
\end{equation}
Note in particular that the equations for the modes are independent of the index $m$.
	
Let us compute the expression of the Lagrangian in terms of the modes. Substituting Eqs. \eqref{defW} and \eqref{defphi} into Eq. \eqref{0.3} and employing the properties of the Fourier modes and of the spherical harmonics, we obtain
\begin{equation} \label{1.10}
\begin{aligned}        
\textbf{L} = \sum_{nlm} \frac{APQ^2}{2} \left[ \frac{ \dot{\phi}_{nlm}^2}{A^2}-W_{nl}  \phi_{nlm}^2\right]. 
\end{aligned}
\end{equation}
	
The canonical momenta of $\phi_{nlm}$ are, in turn, 
\begin{equation}
\Pi_{nlm} = \frac{\delta \textbf{L}}{\delta \dot{\phi}_{nlm}}= \frac{PQ^2}{A} \dot{\phi}_{nlm} .
\end{equation}
With the canonical momentum of the Klein-Gordon field being $\Pi=\frac{\delta \textbf{L}}{\delta \dot{\phi}}=PQ^2 \dot{\phi}/A$, one can check that
\begin{equation}\label{defpi}
\Pi(t,r,\theta,\varphi) =  \sum_{nlm} \Pi_{nlm}(t)  F_n(r) \bar{Y}^m_l(\theta,\varphi) ,
\end{equation}
as expected.
	
Finally, one can obtain the Hamiltonian:
\begin{equation}
\begin{aligned}
\textbf{H} = \sum_{nlm} \Pi_{nlm} \dot{\phi}_{nlm} - \textbf{L} = \frac{A}{PQ^2} \frac{1}{2} \sum_{nlm} \left( \Pi_{nlm}^2 +{P^2Q^4} W_{nl} \phi_{nlm}^2\right).
\end{aligned}
\end{equation}
	
It is clear from the above expressions that, in what concerns the Hamiltonian analysis, there is a privileged choice for the lapse function $A(t)$, namely making it equal to $PQ^2$. This corresponds to the selection of the so-called harmonic time $\tau$, determined by $A(t)dt = P(\tau)Q^2(\tau)d\tau$. We adopt this choice from now on, and we will denote the derivative with respect to $\tau$ by a  prime.
	
\section{New canonical variables}
\label{ncv}
	
In order to address the evolution in the ultraviolet limit, which is pivotal to the issue of unitary dynamics, one would like to decouple the background functions $ P^2$ and $Q^2 $ from the labels $( n , l )$ in the eigenvalues $W_{nl}(\tau)$ that appear in the mode equations. We start by introducing the combined label $k\geq 0$, determined by
\begin{equation}
k= \sqrt{n^2 + l(l+1)}.
\end{equation}
The region of large values of $k$ corresponds precisely to the ultraviolet domain.
	
Leaving the case $k=0$ apart (see our discussion below), let us also introduce the notation $\hat{l}=\sqrt{l(l+1)}/k$. The eigenvalues $W_{nl}$ can now be written as
\begin{equation}
W_{nl} = k^2 \left[\frac{n^2/k^2}{P^2} + \frac{\hat{l}^2}{Q^2}\right]=\frac{k^2}{P^2} \left[1+\hat{l}^2 \left(\frac{ P^2}{Q^2}-1\right)\right] .
\end{equation}
	
Inspired by the previous analysis of unitarity and uniqueness of the Fock quantization in the case of the Bianchi I model  \cite{BI}, we now propose the following time-dependent and mode-dependent canonical transformation:
\begin{equation}\label{z7}
	\begin{pmatrix}
	\tilde{\phi}_{nlm} \\
	\tilde{\Pi}_{nlm}
	\end{pmatrix}
	= \begin{pmatrix}
	\sqrt{b_{\hat{l}}} &  0 \\
	\frac{1}{2} \frac{b^{\prime}_{\hat{l}}}{b_{\hat{l}}^{3/2}} & \frac{1}{\sqrt{b_{\hat{l}}}}
	\end{pmatrix}
	\begin{pmatrix}
	\phi_{nlm} \\
	\Pi_{nlm}
	\end{pmatrix} ,
\end{equation}
where we recall that the prime denotes derivative with respect to $\tau$ and
\begin{equation}
b_{\hat{l}}= Q^2 \sqrt{1 + \hat{l}^2 \left(\frac{P^2}{Q^2}-1\right)}.
\end{equation}
	
Notice that although the mode dependence is necessary to attain unitary quantum dynamics (as mentioned in the Introduction and thoroughly explained in Ref.\ \cite{BI}), this dependence is minimal, since $b_{\hat{l}}$ depends only on the \emph{direction} $\hat{l}$ in the space of labels $( n , l )$, and not on the (almost) \emph{radial} label $k$. Moreover, and most importantly, the proposed canonical transformation fully respects the spatial symmetries -- which are reflected in the spectrum of the LB operator --, in the sense that it does not mix the modes and does not depend on the remaining label $m$. As a technical point to be used further below, let us also mention that $b_{\hat{l}}$ remains bounded both from above and from below when the pair $( n , l )$ runs over all the possible set of values, for any fixed time, assuming just that the metric functions $P$ and $Q$ are continuous.
	
In harmonic time $\tau$, the new Hamiltonian $\tilde{\textbf{H}}$ in terms of the new canonical variables $\tilde{\phi}_{nlm}$ and $\tilde{\Pi}_{nlm}$ is determined by
\begin{equation}
\begin{aligned}
-{\tilde{\sum_{nlm}}} \Pi_{nlm} \phi^{\prime}_{nlm}+ \textbf{H} = -{\tilde{\sum_{n l m}}} \tilde{\Pi}_{nlm} \tilde{\phi}^{\prime}_{nlm} + \tilde{\textbf{H}}.
\end{aligned}
\end{equation}
After a straightforward calculation, we obtain
\begin{equation}
\tilde{\textbf{H}}={\tilde{\sum_{nlm}}} \left[ \frac{b_{\hat{l}}}{2}  \tilde{\Pi}^2_{nlm} + \left(k^2 \frac{b_{\hat{l}}}{2} - \frac{(b^{\prime}_{\hat{l}})^2}{8 b^{3}_{\hat{l}}} \right) \tilde{\phi}_{nlm}^2 + \frac{b_{\hat{l}}^{\prime}}{2b_{\hat{l}}^2} \tilde{\phi}_{nlm} {\tilde{\phi}}_{nlm}^{\prime} \right] .
\end{equation}
Removing from this Hamiltonian a total time derivative, related to the last term in the above expression, we finally get
\begin{equation}\label{z12}
\tilde{\textbf{H}}={\tilde{\sum_{nlm}}}  \frac{b_{\hat{l}}}{2}  \left[\tilde{\Pi}^2_{nlm} + \left(k^2  +\frac{3(b^{\prime}_{\hat{l}})^2}{4b^4_{\hat{l}}} -  \frac{b^{\prime\prime}_{\hat{l}}}{2b^3_{\hat{l}}} \right) \tilde{\phi}_{nlm}^2 \right].
\end{equation}
According to our previous comments, we have left apart the zero mode, corresponding to $n=l=m=0$. The contribution of this mode has been removed (or, rather, viewed as a part of the background), and this is precisely the meaning of the tilde in the new sum $\tilde{\sum}$. This is done strictly to avoid over-complicating the notation. This single mode has no impact whatsoever on the present discussion about unitary implementation of the dynamics. Concerning its quantization (if required),  the zero mode can always be treated separately, by any means available.
	
Once we have reformulated the Hamiltonian description in this manner, we can follow the same steps as in the above-mentioned Bianchi I case, for which the uniqueness of the Fock quantization based on a unitary implementation of the dynamics has been demonstrated \cite{BI}. We present here the main points of this demonstration for the current Kantowski-Sachs case, starting with the analysis of the asymptotic dynamics pertinent to the new canonical variables.
	
\section{Asymptotic dynamics}
\label{acd}
	
The new Hamiltonian equations are 
\begin{equation} \label{1.26}
\tilde{\phi}^{\prime}_{nlm} = b_{\hat{l}}  \tilde{\Pi}_{nlm}, \qquad \tilde{\Pi}^{\prime}_{nlm} = -b_{\hat{l}}\left[ k^2 + s_{\hat{l}}(\tau) \right] \tilde{\phi}_{nlm},
\end{equation}
where 
\begin{equation}
s_{\hat{l}}(\tau)=  \frac{3(b^{\prime}_{\hat{l}})^2}{4b_{\hat{l}}^4} - \frac{ b^{\prime\prime}_{\hat{l}}}{2b_{\hat{l}}^3}.
\end{equation}
Then, we obtain the second-order differential equation:
\begin{equation}\label{0.31}
\frac{1}{b_{\hat{l}}} \frac{d}{d\tau} \left( \frac{1}{b_{\hat{l}}} \frac{d}{d\tau} \tilde{\phi}_{nlm}\right) + (k^2 + s_{\hat{l}}) \tilde{\phi}_{nlm} = 0.
\end{equation}
	
On general grounds, one can always write the solution to Eq.\	\eqref{0.31} in the form 
\begin{equation} \label{0.32}
\tilde{\phi}_{nlm}(\tau) = C_{nlm} e^{i\Theta_{nl}(\tau)} + C_{nlm}^{*} e^{-i\Theta_{nl}^{*}(\tau)},
\end{equation}
where $C_{nlm}$ are complex constants.
	
On the other hand, the dominant behavior of the solution in  the ultraviolet limit $k\to \infty$  is of the form $e^{\pm ik \int  d\tau  b_{\hat{l}}(\tau) }$. In fact, given the definition of $s_{\hat{l}}$ above and the previous comments on the boundedness of $b_{\hat{l}}$, it is easy to see that the absolute value of $s_{\hat{l}}$ remains bounded (if the metric functions are sufficiently smooth and nonvanishing) when the pair $( n , l )$ runs over all the possible set of values, and can therefore be neglected in Eq.\ \eqref{0.31} in the region of large $k$. This argument motivates the following \emph{ansatz} for the functions
$\Theta_{nl}$:  
\begin{equation}\label{1.31}
\Theta_{nl}(\tau) = \int_{\tau_0}^\tau d\bar{\tau} b_{\hat{l}}(\bar{\tau}) \left[ k + i Z_{nl} (\bar{\tau}) \right],
\end{equation}
where $Z_{nl}$ is a subleading correction and $\tau_0$ is some fixed, but arbitrary, initial time. Besides the initial condition for $\Theta_{nl}(\tau_0)$ already implicit  in Eq.\ \eqref{1.31}, we adopt also $Z_{nl}(\tau_0)=0$, giving rise to
\begin{equation}\label{z4}
\begin{aligned}
\Theta_{nl}(\tau_0) = 0, \qquad \Theta^{\prime}_{nl}(\tau_0) = k b_{\hat{l}}(\tau_0).
\end{aligned}
\end{equation}
The first of these conditions is justified by the fact that the initial value of $\Theta_{nl}$ can be absorbed in the definition of the constants $C_{nlm}$. The second condition guarantees that $\Theta_{nl}$ does not vanish in a neighborhood of $\tau_0$ and therefore the exponentials $e^{i  \Theta_{nl}}$ and $e^{- i  \Theta_{nl}^{*}}$ are functionally  independent there for large $k$. 
	
We assume moreover that $\Theta_{nl}$ admits an asymptotic expansion in integer powers of $k$, so that $Z_{nl}$ can be at most of order unity. Let us then study the functions $Z_{nl}$. Substituting Eqs.\ \eqref{0.32} and \eqref{1.31} into Eq.\ \eqref{0.31}, we obtain the following equation:
\begin{equation}\label{z1}
\frac{Z^{\prime}_{nl}}{b_{\hat{l}}} = Z_{nl}^2 - 2ikZ_{nl} + s_{\hat{l}} ,
\end{equation}
which can be linearized in the ultraviolet sector, under the assumption that $Z_{nl} = \mathcal{O}(k^{-1})$ in that regime. This assumption is actually consistent, as we now confirm. Let then  $\tilde{Z}_{nl}$ of order $k^{-1}$ denote the dominant part of ${Z}_{nl}$ in the ultraviolet region. The linearized version of Eq. \eqref{z1} then applies:
\begin{equation}
\begin{aligned}\label{Zeta}
\frac{{\tilde Z}^{\prime}_{nl}}{b_{\hat{l}}}= -2ik \tilde{Z}_{nl} + s_{\hat{l}}. 
\end{aligned}
\end{equation}
This equation is easily solved by introducing the conformal time $\eta_{\hat{l}}$ such that $d\eta_{\hat{l}}=b_{\hat{l}}d\tau$, or
\begin{equation}
\eta_{\hat{l}}(\tau) = \int_{\tau_0}^\tau d\bar{\tau} b_{\hat{l}}(\bar{\tau}), \qquad {\rm with} \quad \eta_{\hat{l}}(\tau_0) = 0 .
\end{equation}
We note that this time depends on the label $\hat{l}$, as does the time-dependent mass term $s_{\hat{l}}(\tau)$ in Eq.\ \eqref{1.26} (these are the main technical differences with respect to previous similar analyses made in Refs.\ \cite{cmsv,cmv2}, in simpler contexts). 
	
Adopting the  initial condition $\tilde{Z}_{nl}(\tau_0)={Z}_{nl}(\tau_0)=0$, we thus get
\begin{equation}\label{z2}
\tilde{Z}_{nl} (\tau) = e^{-2ik\eta_{\hat{l}}(\tau)}  \int_{\tau_0}^\tau d\bar{\tau} b_{\hat{l}}(\bar{\tau})s_{\hat{l}}(\bar{\tau}) e^{2ik\eta_{\hat{l}}(\bar{\tau})},
\end{equation}
or
\begin{equation}
\begin{aligned}
\tilde{Z}_{nl} = -\frac{i}{2k} \left[s_{\hat{l}}(\tau) - s_{\hat{l}}(\tau_0)e^{-2ik\eta_{\hat{l}}(\tau)}- e^{-2ik\eta_{\hat{l}}(\tau)} \int_{\tau_0}^\tau d\bar{\tau}  {s^{\prime}}_{\hat{l}}(\bar{\tau}) e^{2ik\eta_{\hat{l}}(\bar{\tau})} \right] ,
\end{aligned}
\end{equation}
where an integration by parts was performed.
	
Finally, taking the  norm and using the triangle inequality, we obtain
\begin{equation}
\begin{aligned}
\frac{1}{2k} \left|s_{\hat{l}}(\tau) - s_{\hat{l}}(\tau_0) e^{-2ik\eta_{\hat{l}}(\tau)} - e^{-2ik\eta_{\hat{l}}(\tau)} \int^{\tau}_{\tau_0} s^{\prime}_{\hat{l}} e^{2ik\eta_{\hat{l}}} d\bar{\tau}\right|  \leq \frac{1}{2k} \left[|s_{\hat{l}}(\tau)| + |s_{\hat{l}}(\tau_0)| + \int_{\tau_0}^{\tau} |s^{\prime}_{\hat{l}}| d\bar{\tau}\right] =: \frac{C_{\hat{l}}(\tau)}{k} .
\end{aligned}
\end{equation}
Hence, one concludes that there exists  $C_{\hat{l}}$ -- independent of $k$ -- such that $|\tilde{Z}_{nl}| \leq  \frac{C_{\hat{l}}(\tau)}{k} $, confirming that $\tilde{Z}_{nl}$ is indeed of order $k^{-1}$ (under the mild assumption that $s_{\hat{l}}$ is sufficiently smooth as to guarantee  the existence and the continuity of  $C_{\hat{l}}$ as a function of time). It then follows from Eq.\ \eqref{1.31} that $\Theta_{{nl}}$ takes the asymptotic form
\begin{equation}\label{0.38}
\Theta_{{nl}}(\tau) = k \eta_{\hat{l}}(\tau) + \mathcal{O}(k^{-1}),
\end{equation}
for large $k$.
	
The evolution of the momentum variables $\tilde{\Pi}_{nlm}$ follows from the first of the Hamiltonian equations \eqref{1.26}, leading to 
\begin{equation} \label{z3}
\tilde{\Pi}_{nlm}(\tau) = i C_{nlm} \frac{d\Theta_{nl}}{d\eta_{\hat{l}}} e^{i\Theta_{nl}(\tau)} -i C_{nlm}^{*} 
\frac{d\Theta_{nl}^{*}}{d\eta_{\hat{l}}} e^{-i\Theta_{nl}^{*}(\tau)}.
\end{equation}
In particular, one obtains the relation between the constants $C_{nlm}$ and  the initial values $\tilde{\phi}_{nlm}(\tau_0)$ and $\tilde{\Pi}_{nlm} (\tau_0)$
\begin{equation} \label{0.51}
C_{nlm} = \frac{\tilde{\phi}_{nlm}(\tau_0)}{2} - \frac{i\tilde{\Pi}_{nlm} (\tau_0)}{2k},
\end{equation}
taking the second of Eqs.\ \eqref{z4} into account.
	
On the other hand, and following Ref.\ \cite{BI}, to address the ultraviolet limit of the canonical evolution equations, it is more convenient to rewrite  Eq.\ \eqref{z3} in the form
\begin{equation} \label{z5}
\tilde{\Pi}_{nlm}(\tau) = k\left[D_{nlm} e^{i\Xi_{nl}(\tau)} + D_{nlm}^{*} e^{-i\Xi_{nl}^{*}(\tau)}\right],
\end{equation}
where $D_{nlm}$ are constants to be determined by the initial conditions imposed on the new functions $\Xi_{nl}$.
	
Adopting in particular the data
\begin{equation}
\Xi_{nl} (\tau_0) = 0, \qquad \Xi^{\prime}_{nl} = k  b_{\hat{l}}(\tau_0),
\end{equation}
one can check that the  functions $\Xi_{nl}$ again satisfy the asymptotic behavior
\begin{equation} \label{1.42}
\Xi_{nl}(\tau) = k \eta_{\hat{l}}(\tau) + \mathcal{O}(k^{-1}),
\end{equation}
whereas for the constants $D_{nlm}$ we find
\begin{equation} \label{z6}
D_{nlm} = \frac{\tilde{\Pi}_{nlm}(\tau_0)}{2k} + \frac{i}{2} \left[ 1 + \frac{s_{\hat{l}}(\tau_0)}{k^{2}}\right] \tilde{\phi}_{nlm}(\tau_0). 
\end{equation}
	
Finally, Eqs.\ \eqref{0.32} and \eqref{z5} can be rewritten in the standard form of linear evolution equations, from time $\tau_0$ to time $\tau$,
\begin{equation} \label{1.45}
	\begin{pmatrix}
	\tilde{\phi}_{nlm} \\
	\tilde{\Pi}_{nlm}
	\end{pmatrix}_\tau = \mathcal{V}_{nl}(\tau,\tau_0) 
	\begin{pmatrix}
	\tilde{\phi}_{nlm} \\
	\tilde{\Pi}_{nlm}
	\end{pmatrix}_{\tau_0}, \qquad 
	\mathcal{V}_{nl}(\tau,\tau_0) = 
	\begin{pmatrix}
	\mathscr{R} \left\lbrace e^{i\Theta_{nl}} \right\rbrace & \frac{1}{k} \mathscr{I}\left\lbrace e^{i\Theta_{nl}}\right\rbrace \\
	- \frac{k^2+s_{\hat{l}}(\tau_0)}{k}\mathscr{I}\left\lbrace e^{i\Xi_{nl}}\right\rbrace & \mathscr{R} \left\lbrace e^{i\Xi_{nl}} \right\rbrace
	\end{pmatrix} ,
\end{equation}
where the symbols $\mathscr{R}$ and $\mathscr{I}$ denote, respectively, the real and imaginary parts.
	
\section{Fock Quantization with unitary dynamics}
\label{fock}
	
We will now introduce the set of Fock representations of the CCR for the scalar field that are compatible with the spatial symmetries of the Kantowski-Sachs spacetime. Moreover, we will require that those representations support a unitary implementation of the classical field dynamics. We will see that this requirement selects a privileged subclass of representations which, as will be shown in the next section, defines a unique quantization.
	
As mentioned in the Introduction, a particular Fock representation is completely characterized by declaring  which set of classical (complex) variables are to be promoted to creation and annihilation operators in Fock space. Since we want to preserve the spatial symmetry, which in particular translates into dynamical decoupling of the modes and independence with respect to the label $m$, we will restrict our attention to \emph{invariant} Fock representations, such that the annihilation and creation variables are defined by $m$-independent and block diagonal relations of the form\footnote{In fact, the symmetries of the dynamical equations indicate some slightly more restricted relations, with a dependence on $n$ only through its absolute value. We will see that incorporating this additional symmetry is not needed for our uniqueness results.} 
\begin{equation}\label{rep}
	\begin{pmatrix}
	a_{nlm} \\ a_{nlm}^{*}
	\end{pmatrix} = 
	\mathcal{F}_{nl}(\tau)
	\begin{pmatrix}
	\tilde{\phi}_{nlm} \\ \tilde{\Pi}_{nlm}
	\end{pmatrix}, \qquad \text{where} \quad \mathcal{F}_{nl}(\tau) = 
	\begin{pmatrix}
	f_{nl} (\tau)    & g_{nl}(\tau)  \\
	f_{nl}^{*} (\tau) & g_{nl}^{*}(\tau) 
	\end{pmatrix} .
\end{equation}
To ensure that the Poisson brackets between the new variables  $a_{nlm}$ and $a_{nlm}^{*}$ correspond to the standard annihilation-creation algebra, 
one must further impose that
\begin{equation}\label{ccr}
f_{nl}(\tau) g^{*}_{nl}(\tau)  - g_{nl}(\tau) f^{*}_{nl}(\tau) =-i .
\end{equation}
	
Note that we allow time dependence in the definition \eqref{rep}. In this way, we also investigate the possibility of a unitarily implementable (Heisenberg) dynamics different from the one discussed in Secs. \ref{ncv} and \ref{acd}. Putting it differently: the linear transformation defined by Eqs. \eqref{z7} and \eqref{rep} introduces a splitting of the time variation of the field between a part to be considered as assigned to the background, and a part to be  implemented unitarily at the quantum level. The latter is here seen as the genuine quantum evolution. By letting the matrices  $\mathcal{F}_{nl}$ depend on time, we are thus testing the viability of different splittings of the time dependence of the field. The allowed time dependence in Eq. \eqref{rep} is rather arbitrary, restricted only by the obvious condition that one should not completely cancel the dynamical time variation by means of the transformations $\mathcal{F}_{nl}$, which would then simply trivialize the dynamics. We will comment on this condition later in our discussion. 
	
In general, unitary implementation of a given canonical transformation $T$ means that there exists a quantum unitary operator  $\hat T$ that intertwines with the classical action of $T$, thus providing a unitary quantization of the canonical transformation in question. In the case at hand, we require the implementation of the set of canonical transformations corresponding to time evolution, from time $\tau_0$ to time $\tau$, described e.g.\  by the first of Eqs.\ \eqref{1.45}. Concretely, in terms of our new variables, time evolution assumes the form of a set of Bogoliubov transformations
\begin{equation} \label{Bogo}
	\begin{aligned}
	\begin{pmatrix}
	a_{nlm} \\ a_{nlm}^{*}
	\end{pmatrix}_\tau = \mathcal{B}_{nl}(\tau,\tau_0) 
	\begin{pmatrix}
	a_{nlm} \\ a_{nlm}^{*}
	\end{pmatrix}_{\tau_0}, 
	\end{aligned}
\end{equation}
where
\begin{equation}\label{BogoD}
	\begin{aligned}
	\mathcal{B}_{nl}(\tau,\tau_0) = 
	\begin{pmatrix}
	\alpha_{nl} (\tau,\tau_0)   & \beta_{nl}(\tau,\tau_0) \\
	\beta_{nl}^{*}(\tau,\tau_0) & \alpha_{nl}^{*} (\tau,\tau_0) 
	\end{pmatrix}
	= \mathcal{F}_{nl}(\tau)\mathcal{V}_{nl}(\tau,\tau_0)\mathcal{F}^{-1}_{nl}(\tau_0).
	\end{aligned}
\end{equation}
	
Let then $\hat a_{nlm}^{\dagger}$ and  $\hat a_{nlm}$ denote the creation and annihilation operators in Fock space, here seen as the quantum counterparts of the classical variables $a_{nlm}^*(\tau_0)$ and  $a_{nlm}(\tau_0)$, respectively. In other words, we are working in the Fock representation defined by relations \eqref{rep} at time $\tau_0$. Unitary implementation of the time evolution from $\tau_0$ to $\tau$ then means that there exist unitary operators 
$\hat U(\tau,\tau_0)$ such that 
\begin{equation} \label{z8}
\hat U(\tau,\tau_0) \hat a_{nlm} \hat U^{-1}(\tau,\tau_0)= \alpha_{nl} (\tau,\tau_0)\hat a_{nlm} + 
\beta_{nl}(\tau,\tau_0) \hat a_{nlm}^{\dagger},
\end{equation}
and analogously for $\hat a_{nlm}^{\dagger}$. Note that whereas the right-hand side of Eq.\ \eqref{z8} is always well defined, the existence of  $\hat U(\tau,\tau_0)$ is not ensured.  In fact, its existence is tantamount to the unitary equivalence of the two representations defined by the two creation-annihilation pairs, namely the original set and the transformed one, by means of the Bogoliubov transformation \eqref{Bogo}.
	
It follows from general results \cite{Shale} that a necessary and sufficient condition for unitary implementation 
in the present case is
\begin{equation}\label{z9} 
\tilde{\sum_{nlm}}  |\beta_{nl} (\tau,\tau_0)|^2 = \tilde{\sum_{nl}} (2l+1) |\beta_{nl} (\tau,\tau_0)|^2< \infty ,
\end{equation}
for any time $\tau$, where the factor $2l+1$ accounts for the degeneracy (in the label $m$). To discuss the above summability condition, we have to consider the asymptotic behavior of the coefficients $\beta_{nl}$, for large $k$. First, it follows from Eqs. \eqref{1.45} and \eqref{BogoD} that
\begin{equation} \label{2.3}
\begin{aligned}
\beta_{nl}(\tau,\tau_0)  &= -i \left[ f_{nl}(\tau) g_{nl}(\tau_0) \mathscr{R} \left \lbrace e^{i\Theta_{nl}(\tau)} \right \rbrace - g_{nl}(\tau) f_{nl}(\tau_0) \mathscr{R} \left \lbrace e^{i\Xi_{nl}(\tau)} \right \rbrace \right. \\ &\left.  - \frac{1}{k} f_{nl}(\tau)f_{nl}(\tau_0) \mathscr{I} \left \lbrace e^{i\Theta_{nl}(\tau)} \right \rbrace  - \frac{k^2+s_{\hat{l}}(\tau_0)}{k} g_{nl}(\tau) g_{nl} (\tau_0) \mathscr{I}\left \lbrace e^{i\Xi_{nl}(\tau)} \right \rbrace \right] .
\end{aligned}
\end{equation}
Next, our discussion on Sec. \ref{acd} on the asymptotic behavior of the functions $\Theta_{nl}$ and $\Xi_{nl}$ leads to
\begin{equation}\label{z10}
\begin{aligned}
|\beta_{nl}(\tau,\tau_0)| &= \frac{1}{2k} \bigg| \left[f_{nl}(\tau) +ik g_{nl}(\tau) \right] \left[ f_{nl}(\tau_0) - ik g_{nl} (\tau_0) \right] \left[ e^{ik\eta_{\hat l}(\tau)} + \mathcal{O}\left( k^{-1}\right) \right]  \\
&- \left[f_{nl}(\tau) - ikg_{nl}(\tau) \right] \left[ f_{nl} (\tau_0) +ikg_{nl}(\tau_0) \right] 
\left[ e^{-ik\eta_{\hat l}(\tau)} + \mathcal{O}\left(k^{-1}\right) \right] \\ &+2i  g_{nl}(\tau)g_{nl}(\tau_0) s_{\hat{l}}(\tau_0) \left[ 
\sin(k \eta_{\hat l}(\tau)) + \mathcal{O} \left( k^{-1} \right) \right] \bigg| .
\end{aligned}
\end{equation}
	
We are now in a position to deduce the conditions imposed on the matrices $\mathcal{F}_{nl}$ by the unitary dynamics requirement \eqref{z9}. We follow here the line of reasoning presented in Ref.\ \cite{BI}. Thus, in particular, we do not allow for matrices $\mathcal{F}_{nl}$ that could absorb in their time dependence the dominant dynamical behavior of the field modes, given by the phases $\exp(\pm ik\eta_{\hat{l}})$. This would lead to an undesired trivialization of the dynamics, in line with our comments at the beginning of this section. Keeping this restriction in mind, let us start by noting that the last term in Eq.\ \eqref{z10} is subdominant compared to other contributions proportional to $g_{nl}(\tau)g_{nl}(\tau_0)$ present in the first two terms. Assuming continuity in time of $\mathcal{F}_{nl}$, one can then show that the first two  terms on the right-hand side of Eq.\ \eqref{z10} must satisfy the  summability condition independently. This, together with Eq.\ \eqref{ccr}, can be seen to determine the functions $f_{nl}$ and $g_{nl}$ at leading order to be
\begin{equation} \label{z11}
\breve{f}_{nl}(\tau) = \sqrt{\frac{k}{2}} , \qquad \breve{g}_{nl}(\tau) = \frac{i}{\sqrt{2k}}.
\end{equation}
We notice that, in principle, both $\breve{f}_{nl}$  and $\breve{g}_{nl}$ could be multiplied by a common phase $e^{iG_{nl}}$, that could be made time dependent. However, those phases play no role in the determination of the Fock representation and we therefore obviate them.  
	
At this point, one can  go back to Eq.\ \eqref{z10} and check that the third term satisfies the summability condition by itself, as we now show. 
Given that $s_{\hat l}$ was seen to remain bounded on the space of values $(n,l)$ (and the same obviously happens with the sinus function), the relevant part of the sum in question is bounded by one proportional to
\begin{equation}
\sum_{nl} (2l+1)\frac{1}{k^4},
\end{equation}
which converges to a finite value. In fact, the number of values of $l$ for a given $k$ grows asymptotically like $k$, because we can think of $k$ as the radius of a semicircle in a half-plane ($l \geq 0$) in the considered asymptotic limit. It follows that the above sum can be reexpressed as a sum over $k$ with a summand that is bounded by $1/k^2$ (modulo a constant), and therefore converges.
	
Although fixing $f_{nl}$ and $g_{nl}$ by expressions \eqref{z11} is sufficient to satisfy the unitary dynamics condition \eqref{z9}, unitarity allows more freedom for these functions. Let us then generalize expressions \eqref{z11} by including subdominant terms in the form 
\begin{equation} \label{2.7}
f_{nl}(\tau) = \sqrt{\frac{k}{2}} 
+ k \vartheta_{nl}^{(f)}(\tau), \qquad g_{nl}(\tau) = \frac{i}{\sqrt{2k}} 
+ \vartheta_{nl}^{(g)}(\tau).
\end{equation}
Note first that relation \eqref{ccr} imposes restrictions on the subdominant contributions, namely
\begin{equation} \label{2.9}
\mathscr{R} \left \lbrace \vartheta_{nl}^{(f)}  \right \rbrace +
\mathscr{I} \left \lbrace   \vartheta_{nl}^{(g)} \right \rbrace + \sqrt{2k}\mathscr{I} \left \lbrace    \vartheta_{nl}^{(g)}  \left[\vartheta_{nl}^{(f)}\right]^{*}\right \rbrace 
=0.
\end{equation}
On the other hand, the summability of the contributions to the first two terms in Eq.\ \eqref{z10} coming from the subdominant terms amounts to
\begin{equation} \label{2.8}
\tilde{\sum_{nlm}} k | \vartheta_{nl}^{(f)}(\tau) + i \vartheta_{nl}^{(g)}(\tau) |^2 = \tilde{\sum_{nl}} (2l+1)k | \vartheta_{nl}^{(f)}(\tau) + i \vartheta_{nl}^{(g)}(\tau) |^2< \infty 
\end{equation}
at all $\tau$. This last condition, together with the general form \eqref{2.7} and the relations \eqref{2.9}, completely characterizes the subclass of invariant Fock representations giving rise to a unitary implementation of the (Heisenberg) dynamics. By the same token, the possible time-dependent contributions to relations \eqref{rep} are severely restricted.
	
\section{Uniqueness of the Quantization}
\label{uni}
	
As we have seen, invariant Fock representations for the scalar field,  i.e.\ those compatible with the spatial symmetries, are characterized by a particular set of (possibly time-dependent) matrices $\mathcal{F}_{nl}$ \eqref{rep} (see also our comments in the footnote). From those representations, the requirement of unitary implementation of the dynamics selects a special subclass. We will now show that all the elements of this subclass are unitarily equivalent, i.e.\  there is a unitary transformation relating any given two, and in this sense the quantization emerging from our requirements is unique. 
	
We start by  choosing a reference representation in the subclass, to which any other can be easily related. Naturally, for simplicity, we choose the representation determined by Eqs. \eqref {z11}, and we adopt the symbol ${}\breve{ }$ (`breve') to refer to quantities corresponding to this reference representation.
	
Let then  $\mathcal{F}_{nl}$   be a set of matrices associated with an invariant Fock  representation as in 
Eq. \eqref{rep}.  At any given time $\tau$, the corresponding pairs of creation and annihilation variables are then related to the reference ones by the following Bogoliubov transformation:
\begin{equation}
	\begin{pmatrix}
	a_{nlm}\\
	a_{nlm}^{*}
	\end{pmatrix}_\tau =  
	\mathcal{K}_{nl}(\tau) \begin{pmatrix}
	\breve{a}_{nlm}\\
	\breve{a}_{nlm}^{*}
	\end{pmatrix}_\tau , \qquad \mathcal{K}_{nl} (\tau)= \mathcal{F}_{nl}(\tau)  \breve{\mathcal{F}}^{-1}_{nl}(\tau) =
	\begin{pmatrix}
	\kappa_{nl} (\tau)& \lambda_{nl}(\tau)\\
	\lambda_{nl}^{*} (\tau)& \kappa_{nl}^{*}(\tau)
	\end{pmatrix}.
\end{equation}
The above functions $\kappa_{nl}$ and $\lambda_{nl}$ are easily obtained. In particular, we have 
\begin{equation}
\lambda_{nl}(\tau) =-i \left[ f_{nl}(\tau) \breve{g}_{nl}(\tau) - g_{nl}(\tau)\breve{f}_{nl}(\tau)\right].
\end{equation}
	
Again, the two representations, given respectively by $\mathcal{F}_{nl}$ and $\breve{\mathcal{F}}_{nl}$, are unitarily equivalent if and only if 
\begin{equation}
{\tilde{\sum}}_{nlm}|\lambda_{nl}(\tau)|^2<\infty.
\end{equation}
Finally, if $\mathcal{F}_{nl}$ is of the form determined by Eqs.\ \ref{2.7}, we obtain simply  
\begin{equation}
\lambda_{nl}(\tau) =\sqrt{\frac{k}{2}}\left(\vartheta_{nl}^{(f)}(\tau) + i\vartheta_{nl}^{(g)}(\tau)\right).
\end{equation}
Thus, the condition for unitary equivalence of the two representations coincides with condition \eqref{2.8}, and it is therefore guaranteed to be fulfilled if the second representation also belongs to the subclass selected by the unitary dynamics requirement. Since unitary equivalence is transitive, it follows that all elements of the subclass are unitarily equivalent to each other.
	
Let us discuss the results of these last two sections, starting by restricting attention to the situation where the matrices $\mathcal{F}_{nl}$ are taken to be time-independent. In that case, the time-dependence in the Bogoliubov transformations \eqref{Bogo} is strictly the one coming from the classical dynamics of the canonical system $(\tilde \phi_{nlm}, \tilde \Pi_{nlm})$ defined by Eqs.\ \eqref{z7}. Relations \eqref{rep} are then simply parametrizing the set of invariant Fock representations (at fixed time). Among those representations, the requirement of unitary implementation of the dynamics, generated by the Hamiltonian \eqref{z12},  selects a special subclass, of the form \eqref{2.7}, with the (time-independent) coefficients $\vartheta_{nl}^{(f)}$ and $\vartheta_{nl}^{(g)}$ restricted by conditions \eqref{2.9} and \eqref{2.8}.  Thus, there is a Fock quantization with unitary dynamics for the system $(\tilde \phi_{nlm}, \tilde \Pi_{nlm})$. Furthermore,  this quantization is unique, since it follows immediately from the results of the current section that all the selected representations are unitarily equivalent. The canonical transformation \eqref{z7} therefore succeeds in splitting the time evolution of the original canonical pair $(\phi,\Pi)$ such that a unique quantization with unitary dynamics is achieved. This replicates previous results obtained in isotropic and conformally ultrastatic scenarios, and already gives a privileged status to the canonical transformation \eqref{z7}.
	
A second point is whether the dynamics, or rather the splitting between background and field dynamics, is uniquely fixed. This happens e.g.\ in previously analyzed situations \cite{cmov2,cmov4,cmvT} where a global (i.e.\ mode-independent) scaling is sufficient to achieve unitary dynamics.  In that case, there is no reason to allow mode-dependent (and time-dependent) transformations, and additional mode-independent transformations are incompatible with unitary dynamics, because they affect equally the whole set of modes \cite{cmvT}. In the present case, because the canonical transformation \eqref{z7} is itself mode dependent, we allowed mode-dependent variations, and alternative dynamics compatible with unitary implementation were bound to appear. In particular, changing the dynamics on a finite number of modes certainly has no impact on the unitary implementation. More generally, we have found that our criteria fix the dynamics of the annihilation and creation variables at the dominant order, allowing only extra subdominant time-dependent contributions, which are required to satisfy the condition \eqref{2.8}.
	
In any case, note that  the subdominant alternative dynamics do not require new representations. In fact, what was shown in Sec. \ref{fock} is that any of the allowed choices of dynamics is unitarily implementable with respect to the Fock representation defined by relations \eqref{rep} at time $\tau_0$. But again that representation is unitarily equivalent to our reference representation, and it therefore follows that all admissible dynamics are unitarily implementable in the same Fock representation.
	
\section{Conclusions}
\label{con}
	
We have considered a Klein-Gordon field in a Kantowski-Sachs spacetime and studied criteria for the choice of a unique family of unitarily equivalent Fock representations for its quantization. The criteria consist of the invariance of the representation under the spatial isometries of the background (or, similarly, of the dynamical equations for the field propagating on it) and in the unitary implementation of the Heisenberg evolution of the creation and annihilation operators of the representation. These criteria had already been proven successful in selecting a unique Fock representation (up to unitary transformations) in a series of cosmological scenarios, including some anisotropic ones. Our extension of the uniqueness result to Kantowski-Sachs reinforces the robustness of these criteria, showing that they remain valid in much more general cases than conformally ultrastatic spacetimes where the field can be locally redefined by a scaling that leads to dynamics with good ultraviolet properties. Furthermore, the case of Kantowski-Sachs opens potential applications to black hole physics, because the interior geometry of a nonrotating black hole is isometric to a cosmological spacetime of this kind. In this way, we have a procedure at hand to choose a family of Fock representations for fields in the interior of black holes and construct quantum (Heisenberg) unitary dynamics for them. Extensions to the exterior of both the black hole geometry and the constructed Fock quantum field theory can provide valuable tools to understand the properties of the vacuum and notions of unitarity for an exterior observer. 
	
It is worth emphasizing that our criteria only select a family of Fock representations that are all mutually equivalent. The determination of a vacuum state for the quantum field theory in the interior still needs further restriction in the choice of representation until no ambiguity is left. In terms of the transformations analyzed in our work, this ambiguity is the remaining freedom in the specification of subdominant terms in Eq. \eqref{2.7}. A possibility that has been recently put forward is fixing them by demanding an asymptotic diagonalization in the ultraviolet of the generator of the Heisenberg dynamics \cite{EMT}. The vacuum picked out in this manner has been argued to display a power spectrum with remarkable non-oscillating behavior, at least for certain effective backgrounds that appear in quantum cosmology \cite{NO}. It would be very interesting to study if this proposal can be applied in Kantowski-Sachs and determine the corresponding vacuum state for the Klein-Gordon field. Extension of the arguments to the exterior would allow us to discuss the relationship of this vacuum with other vacua suggested for scalar fields in black hole spacetimes \cite{wald}.
	
As we have discussed at the end of the previous section, two important ingredients in our treatment are the time dependence and the mode dependence that we have allowed in our canonical transformations. By the combination of these two ingredients, we have been able to characterize a unique family of dynamics that are unitarily implementable. The evolution of the Klein-Gordon field is the compound of an explicit dependence on the time-varying background, on the one hand, and of the quantum evolution of the annihilation and creation variables, on the other hand. It is this last quantum dynamical component that can be promoted to a unitary transformation in our class of Fock representations. All these representations have been shown to be unitarily equivalent. Moreover, all the dynamical transformations that belong to the above unitarily implementable family are related by unitary operators among them, in any of the admissible Fock representations. In this sense, our criteria are able to extract a part of the dynamics that can be implemented as a unitary for Kantowski-Sachs. Again, an extension of the geometry and the quantum field theory to the exterior would allow us to discuss the implications of our dynamical splitting for probability conservation, investigating which part of the evolution would preserve it and what is the notion of unitary transformation that one would reach.   
	
\acknowledgments
	
This work was partially supported by Project No. MICINN PID2020-118159GB-C41 from Spain, Grants NSF-PHY-1903799, NSF-PHY-2206557, and funds of the Hearne Institute for Theoretical Physics. J.M.V.\ is grateful for the support given by the research unit Fiber Materials and Environmental Technologies (FibEnTech-UBI), on the extent of the project reference UIDB/00195/2020, funded by the Funda\c c\~ao para a Ci\^encia e a Tecnologia (FCT), IP/MCTES through national funds (PIDDAC). The authors are grateful to A. Garc\'{\i}a-Quismondo for enlightening discussions.


\begin{thebibliography}{99.}
	
\bibitem{KS1} R. Kantowski, and R.K. Sachs, Some spatially inhomogeneous dust models, J. Math. Phys. {\bf 7}, 443 
	(1966).
	
\bibitem{KS2} E. Weber, Kantowski–Sachs cosmological models as big-bang models, J. Math. Phys. {\bf 26}, 1308 (1985).
	
\bibitem{KS3} K.S. Adhav, V.G. Mete, A.S. Nimkar, and A.M. Pund, Kantowski-Sachs cosmological model in general theory of relativity, Int. J. Theor. Phys. {\bf 47}, 2314 (2008).
	
\bibitem{AB} A. Ashtekar, and M. Bojowald, Quantum geometry and the Schwarzschild singularity, Class. Quantum Grav. {\bf 23}, 391 (2006).
	
\bibitem{ALQG} A. Ashtekar and J. Lewandowski, Background independent quantum gravity: A status report,  Class. Quantum Grav. \textbf{21}, R53 (2004).
	
\bibitem{AOS} A. Ashtekar, J. Olmedo, and P. Singh, Quantum transfiguration of Kruskal black holes, Phys. Rev. Lett. {\bf 121}, 241301 (2018). 
	
\bibitem{AOS2} A. Ashtekar, J. Olmedo, and P. Singh, Quantum extension of the Kruskal spacetime, Phys. Rev. D {\bf 98}, 126003 (2018).
	
\bibitem{AO} A. Ashtekar and J. Olmedo, Properties of a recent quantum extension of the Kruskal geometry, Int. J. Mod. Phys. D {\bf 29}, 2050076 (2020).
	
\bibitem{EGM} B. Elizaga Navascu\'es, A. Garc\'{\i}a-Quismondo, and G.A. Mena Marug\'an, Hamiltonian formulation and loop quantization of a recent extension of the Kruskal spacetime, Phys. Rev. D {\bf 106}, 043531 (2022).
	
\bibitem{EMM} B. Elizaga Navascu\'es, G.A. Mena Marug\'an, and A. M\'{\i}nguez-S\'anchez, Extended phase space quantization of a black hole interior model in loop quantum cosmology, arXiv:2306.06090 (2023).
	
\bibitem{hLQC} B. Elizaga Navascu\'es and G.A. Mena Marug\'an,  Hybrid loop quantum cosmology: An overview, Front. Astron. Space Sci. {\bf 8}, 624824 (2021).
	
\bibitem{FMO} M. Fern\'andez-M\'endez, G.A. Mena Marug\'an, and J. Olmedo, Hybrid quantization of an inflationary model: The flat case, Phys. Rev. D {\bf 88}, 044013 (2013).
	
\bibitem{CMM} L. Castell\'o Gomar, M. Mart\'{\i}n-Benito, and G.A. Mena Marug\'an, Gauge-invariant perturbations in hybrid quantum cosmology, JCAP {\bf 06} (2015) 045.
	
\bibitem{wald} R.M. Wald,  \emph{Quantum Field Theory in Curved Spacetime and Black Hole Thermodynamics} (Chicago University Press, Chicago, 1994).
	
\bibitem{kayb} B. Kay,  Linear spin-zero quantum fields in external gravitational and scalar fields, Commun. Math. Phys. {\bf 62}, 55 (1978).
	
\bibitem{ccmv} A. Corichi, J. Cortez, G.A. Mena Marug\'an, and J.M. Velhinho,  Quantum Gowdy $T^3$ model: A uniqueness result, Class. Quantum Grav. {\bf 23}, 6301 (2006).
	
\bibitem{cmov2} J. Cortez, G.A. Mena Marug\'an, J. Olmedo, and J.M. Velhinho,  Uniqueness of the Fock quantization of fields with unitary dynamics in nonstationary spacetimes, Phys. Rev. D {\bf 83}, 025002 (2011).
	
\bibitem{cmov3} J. Cortez, G.A. Mena Marug\'an, J. Olmedo, and J.M. Velhinho,  A uniqueness criterion for the Fock quantization of scalar fields with time dependent mass, Class. Quantum Grav. {\bf 28}, 172001 (2011).
	
\bibitem{cmov4} J. Cortez, G.A. Mena Marug\'an, J. Olmedo, and J.M. Velhinho,  Criteria for the determination of time dependent scalings in the Fock quantization of scalar fields with a time dependent mass in ultrastatic spacetimes, Phys. Rev. D {\bf 86}, 104003 (2012).
	
\bibitem{fmov} M. Fern\'andez-M\'endez, G.A. Mena Marug\'an, J. Olmedo, and J.M. Velhinho,  Unique Fock quantization of scalar cosmological perturbations, Phys. Rev. D {\bf 85}, 103525 (2012).
	
\bibitem{CJCAP} L. Castell\'o  Gomar, J. Cortez, D. Mart\'{\i}n-de Blas, G.A. Mena Marug\'an, and J.M. Velhinho, Uniqueness of the Fock quantization of scalar fields in spatially flat cosmological spacetimes, JCAP {\bf 11} (2012) 001.
	
\bibitem{cmv} J. Cortez, G.A. Mena Marug\'an, and J.M. Velhinho,  Quantum unitary dynamics in cosmological spacetimes, Ann. Phys. {\bf 363}, 36 (2015).
	
\bibitem{cmv2} J. Cortez, G.A. Mena Marug\'an, and J.M. Velhinho, A brief overview of results about uniqueness of the quantization in cosmology, Universe {\bf 7}, 299 (2021).  
	
\bibitem{cmvT} J. Cortez, G.A. Mena Marug\'an, and J.M. Velhinho,  Uniqueness of the Fock quantization of the Gowdy $T^3$ model, Phys. Rev. D {\bf 75}, 084027 (2007).
	
\bibitem{cmsv} J. Cortez, G.A. Mena Marug\'an, R. Ser\^odio, and J.M. Velhinho,  Uniqueness of the Fock quantization of a free scalar field on $S^1$ with time dependent mass, Phys. Rev. D {\bf 79}, 084040 (2009).
	
\bibitem{CM} L. Castell\'o  Gomar and G.A. Mena Marug\'an,  Uniqueness of the Fock quantization of scalar fields and processes with signature change in cosmology, Phys. Rev. D {\bf 89}, 084052 (2014).
	
\bibitem{cmvS} J. Cortez, G.A. Mena Marug\'an, and J.M. Velhinho, Fock quantization of a scalar field with time dependent mass on the three-sphere: Unitarity and uniqueness, Phys. Rev. D {\bf 81}, 044037 (2010).
	
\bibitem{fermi} J. Cortez, B. Elizaga Navascu\'es, M. Mart\'{\i}n-Benito, G.A. Mena Marug\'an, and J.M. Velhinho, Unique Fock quantization of a massive fermion field in a cosmological scenario, Phys. Rev. D {\bf 93}, 084053 (2016).
	
\bibitem{fermi2} J. Cortez, B. Elizaga Navascu\'es, M. Mart\'{\i}n-Benito, G.A. Mena Marug\'an, and J.M. Velhinho, Uniqueness of the Fock quantization of Dirac fields in 2+1 dimensions, Phys. Rev. D {\bf 96}, 025024 (2017).
	
\bibitem{fermi3} J. Cortez, B. Elizaga Navascu\'es, M. Mart\'{\i}n-Benito, G.A. Mena Marug\'an, and J.M. Velhinho, Dirac fields in flat FLRW cosmology: Uniqueness of the Fock quantization, Ann. Phys. {\bf 376}, 76 (2017).
	
\bibitem{fermi4} J. Cortez, B. Elizaga Navascu\'es, G.A. Mena Marug\'an, S. Prado, and J.M. Velhinho, Uniqueness criteria for the Fock quantization of Dirac fields and applications in hybrid loop quantum cosmology, Universe {\bf 6}, 241 (2020).
	
\bibitem{BI} J. Cortez, B. Elizaga Navascu\'es, M. Mart\'{\i}n-Benito, G.A. Mena Marug\'an, J. Olmedo, and J. M. Velhinho, Uniqueness of the Fock quantization of scalar fields in a Bianchi I cosmology with unitary dynamics, Phys. Rev. D {\bf 94}, 105019 (2016).
	
\bibitem{fon} J. Cortez, L. Fonseca, D. Mart\'{\i}n-de Blas, and G.A. Mena Marug\'an,  Uniqueness of the Fock quantization of scalar fields under mode preserving canonical transformations varying in time, Phys. Rev. D {\bf 87}, 044013 (2013).
	
\bibitem{KSBH1}  J.J. Halliwell and J. Louko, Steepest-descent contours in the path-integral approach to quantum cosmology. III. A general method with applications to anisotropic minisuperspace models, Phys. Rev. D {\bf 42}, 3997 (1990).
	
\bibitem{KSBH2} K.V. Kucha\v{r}, Geometrodynamics of Schwarzschild black holes, Phys. Rev. D {\bf 50}, 3961 (1994).
	
\bibitem{KSBH3} R. Gambini and J. Pullin, Loop quantization of the Schwarzschild black hole, Phys. Rev. Lett. {\bf 110}, 211301 (2013).
	
\bibitem{KSBH4} A. Alonso-Serrano, L.J. Garay, and G.A. Mena Marug\'an, Correlations across horizons in quantum cosmology, Phys. Rev. D {\bf 90}, 124074 (2014).
	
\bibitem{KSBH5} F. Mercati and D. Sloan, Traversing through a black hole singularity, Phys. Rev. D {\bf 106}, 044015 (2022).
	
\bibitem{KS-SF} B.C. Xanthopoulos and T. Zannias, Kantowski–Sachs metrics with source: A massless scalar field, J. Math. Phys. {\bf 33}, 1415. (1992).
	
\bibitem{Shale} D. Shale,  Linear symmetries of free boson fields, Trans. Am. Math. Soc. {\bf 103}, 149 (1962).
	
\bibitem{EMT} B. Elizaga Navascu\'es, G.A. Mena Marug\'an, and T. Thiemann, Hamiltonian diagonalization in hybrid quantum cosmology, Class. Quantum Grav. {\bf 36}, 18 (2019).
	
\bibitem{NO} B. Elizaga Navascu\'es, G.A. Mena Marug\'an, and S. Prado, Non-oscillating power spectra in loop quantum cosmology, Class. Quantum Grav. {\bf 38}, 035001 (2020).
	
\end{thebibliography}
\end{document}